\documentclass[12pt]{amsart}

\usepackage{amsmath,amsthm,txfonts}
 
\newcommand{\pa}{\partial}
\newcommand{\opa}{\overline{\partial}}
\newcommand{\ok}{\overline{k}}

\renewcommand{\Im}{\operatorname{Im}}
\newcommand{\Res}{\operatorname{Res}}

\newcommand{\RR}{\mathbb{R}}

\newcommand{\ga}{\gamma}
\newcommand{\de}{\delta}
\newcommand{\ka}{\kappa}

\newcommand{\tchi}{\widetilde{\chi}}
\newcommand{\trho}{\widetilde{\rho}}
\newcommand{\tQ}{\widetilde{Q}}
\newcommand{\tc}{\widetilde{c}}
\newcommand{\tr}{\widetilde{r}}
\newcommand{\tT}{\widetilde{T}}
\newcommand{\tka}{\widetilde{\kappa}}

\theoremstyle{plain}
\newtheorem{theorem}{Theorem}

\newtheorem{proposition}[theorem]{Proposition}

\theoremstyle{definition}

\usepackage{color}

\begin{document}

\title{Generalized primitive potentials}

\author[D.~Zakharov]{Dmitry Zakharov}
\address{Department of Mathematics, Central Michigan University, Mount Pleasant, MI}
\email{dvzakharov@gmail.com}
\author[V.~Zakharov]{Vladimir Zakharov}
\address{Skolkovo Institute of Science and Technology, Moscow, Russia}
\address{Department of Mathematics, University of Arizona, Tucson, Arizona, USA}
\email{zakharov@math.arizona.edu}

\begin{abstract} In our previous work, we introduced a new class of bounded potentials of the one-dimensional Schr\"odinger operator on the real axis, and a corresponding family of solutions of the KdV hierarchy. These potentials, which we call primitive, are obtained as limits of rapidly decreasing reflectionless potentials, or multisoliton solutions of KdV. In this note, we introduce generalized primitive potentials, which are obtained as limits of all rapidly decreasing potentials of the Schr\"odinger operator. These potentials are constructed by solving a contour problem, and are determined by a pair of positive functions on a finite interval and a functional parameter on the real axis. 
\begin{description}
\item[Keywords] integrable systems, Schr\"odinger equation, primitive potentials
\end{description}

\end{abstract}

\maketitle

\section{Introduction}

This paper is concerned with the spectral properties of the one-dimensional Schr\"odinger operator on the real axis:
\begin{equation}
-\psi''+u(x)\psi=E\psi,\quad -\infty<x<\infty.
\label{eq:Sch}
\end{equation}
The Schr\"odinger operator is the auxiliary linear operator for the KdV hierarchy, the first equation of which is 
\begin{equation}
u_t(x,t)=6u(x,t)u_x(x,t)-u_{xxx}(x,t).
\label{eq:KdV}
\end{equation}
The intimate relationship between the Schr\"odinger operator and KdV is one of the starting points for the modern theory of integrable systems. 

The KdV equation can be solved exactly using the Inverse Scattering Method (ISM) if the initial condition $u(x)=u(x,0)$ is rapidly decreasing as $x\to \pm \infty$. The Schr\"odinger operator~\eqref{eq:Sch} with a rapidly decreasing $u(x)$ has a finite number of bound states with negative energy, and a doubly degenerate spectrum for $E>0$ having, in general, a nontrivial reflection coefficient. The ISM constructs an auxiliary function $\chi(k,x)$ depending on a spectral parameter $k=\sqrt{E}$. This function satisfies a nonlocal Riemann--Hilbert problem, and has the following singularities in the $k$-plane: $N$ poles on the positive imaginary axis, corresponding to the bound states (i.e.~to KdV solitons), and a jump on the real axis determined by the reflection coefficient. If $u(x,t)$ varies according to KdV with initial condition $u(x,0)=u(x)$, the residues and jumps of $\chi(k,x,t)$ satisfy linear ODEs and vary exponentially, and the asymptotic behavior of $\chi(k,x,t)$ as $|k|\to \infty$ determines $u(x,t)$. When the reflection coefficient is equal to zero, we obtain Bargmann's reflectionless potentials of the Schr\"odinger operator and $N$-soliton solutions of KdV. 

There exists another method for constructing solutions of KdV, known as the finite-gap method. The resulting solutions $u(x,t)$ are quasiperiodic in $x$, and hence not rapidly decreasing at infinity. The spectrum of the corresponding Schr\"odinger operator consists of finitely many allowed bands, in which the potential is reflectionless, separated by forbidden gaps. The corresponding function $\chi$ has jumps along the gaps, and can be viewed as a meromorphic function on a hyperelliptic Riemann surface that is a double cover of the $k$-plane branched along the gaps. It is a folklore result that finite-gap solutions should arise from $N$-soliton solutions in the $N\to\infty$ limit, but a precise description of such a limit was lacking. 

Our previous papers~\cite{DZZ16,ZDZ16,ZZD16,ZZ18,NZZ18} offered a solution to this problem, consisting of two steps. Consider the family of $N$-soliton solutions of KdV. The corresponding function $\chi$ is meromorphic on the $k$-plane with $N$ poles on the positive imaginary axis, and its residues satisfy a linear system of equations with $N$ free parameters. First, we allow $\chi$ to have poles on the negative as well as the positive imaginary axis. We then take the limit $N\to\infty$ and obtain a function $\chi$ having jumps on two symmetric cuts $[ia,ib]$ and $[-ib,-ia]$ on the imaginary axis. These jumps satisfy a pair of singular integral equations determined by two arbitrary positive functions $R_1$ and $R_2$ on $[a,b]$, called the {\it dressing functions}. The resulting {\it primitive potentials} of the Schr\"odinger operator have a doubly degenerate spectrum $[-b^2,-a^2]\cup [0,\infty)$ and are reflectionless on $[0,\infty)$, while the corresponding solutions of KdV can be viewed as a dense soliton gas. The elliptic one-gap potential is obtained by taking $R_1=R_2=1$. More generally, Nabelek shows in~\cite{N09} that all finite-gap potentials can be obtained in this way, by setting $R_1=R_2$ to be the indicator function of the set of finite allowed bands, and then varying along a higher KdV flow.

In this note, we perform the same limiting procedure, but for a general rapidly decreasing potential of the Schr\"odinger operator, having $N$ bound states and a nontrivial reflection coefficient. In the $N\to \infty$ limit, the auxiliary function $\chi(k,x)$ has jumps on two symmetric cuts on the imaginary axis, corresponding to a soliton gas, and a jump along the real axis determined by the reflection coefficient. The corresponding potentials of the Schr\"odinger operator, which we call {\it generalized primitive potentials}, have the same doubly degenerate spectrum $[-b^2,-a^2]\cup [0,\infty)$, but are no longer reflectionless on $[0,\infty)$. The function $\chi$ is obtained by solving a Riemann--Hilbert problem involving two arbitrary positive functions $R_1$ and $R_2$ on $[a,b]$, and a functional parameter $c(k)$ on the real axis, inheriting the properties of the reflection coefficient in the rapidly decreasing case.

\section{The inverse scattering transform as an $\opa$-problem} \label{sec:IST}

We begin by recalling the inverse scattering method (see \cite{NMPZ80,DT79,GT09}), and its reformulation as an $\opa$-problem. Consider the Schr\"odinger operator
\begin{equation}
L(t)=-\frac{d^2}{dx^2}+u(x,t)
\end{equation}
with $u(x,t)$ satisfying
\begin{equation}
\int_{-\infty}^{\infty}u(x,0)(1+|x|)dx<\infty \label{eq:udecay}
\end{equation}
and the KdV equation \eqref{eq:KdV}. The operator $L(t)$ has a finite number of bound states with eigenvalues $-\ka_1^2,\ldots,-\ka_N^2$ not depending on $t$, and has an absolutely continuous spectrum on $[0,\infty)$. The Schr\"odinger equation admits two Jost solutions
\begin{equation}
L(t)\psi_{\pm}(k,x,t)=k^2\psi_{\pm}(k,x,t),\quad \Im(k)>0, \label{eq:Jost}
\end{equation}
with asymptotic behavior
\begin{equation}
\lim_{x\to \pm\infty} e^{\mp ikx}\psi_{\pm}(k,x,t)=1.
\end{equation}
The Jost solutions $\psi_{\pm}$ are analytic for $\Im k>0$ and continuous for $\Im k\geq 0$, and have the following asymptotic behavior as $k\to \infty$ with $\Im k>0$:
\begin{equation}
\psi_{\pm}(k,x,t)=e^{\pm ikx}\left(1+Q_{\pm}(x,t)\frac{1}{2ik}+O\left(\frac{1}{k^2}\right)\right),
\end{equation}
where
\begin{equation}
Q_+(x,t)=-\int_x^{\infty} u(y,t)\, dy,\quad Q_-(x,t)=-\int_{-\infty}^x u(y,t)\, dy.
\end{equation}
The Jost solutions satisfy the scattering relations
\begin{equation}
t(k)\psi_{\mp}(k,x,t)=\overline{\psi_{\pm}(k,x,t)}+r_{\pm}(k,t)\psi_{\pm}(k,x,t),\quad k\in \RR,
\end{equation}
where $t(k)$ and $r_{\pm}(k,t)$ are the transmission and reflection coefficients, respectively. These coefficients satisfy the following properties:

\begin{proposition} The transmission coefficient $t(k)$ is meromorphic for $\Im k>0$ and is continuous for $\Im k\geq 0$. It has simple poles at $i\ka_1,\ldots,i\ka_N$ with residues
\begin{equation}
\Res_{i\ka_n} t(k)=i\mu_n(t)\ga_n (t)^2,
\end{equation}
where
\begin{equation}
\ga_n(t)^{-1}=||\psi_+(i\ka_n,x,t)||_2,\quad \psi_+(i\ka_n,x,t)=\mu_n(t)\psi_-(i\ka_n,x,t).
\end{equation}
Furthermore,
\begin{equation}
t(k)\overline{r_+(k,t)}+\overline{t(k)}r_-(k,t)=0,\quad |t(k)|^2+|r_{\pm}(k,t)|^2=1,\quad r_-(k,t)=\overline{r}_+(k,t).
\end{equation}
If we denote $r(k,t)=r_+(k,t)$, $r(k)=r(k,0)$, and $\ga_n=\ga_n(0)$, then
\begin{equation}
t(-k)=\overline{t(k)},\quad r(-k)=\overline{r(k)},\quad k\in \RR, \label{eq:symmetryofreflection}
\end{equation}
\begin{equation}
|r(k)|<1\mbox{ for }k\in \RR,k\neq 0,\quad r(0)=-1\mbox{ if }|r(0)|=1,\label{eq:propertiesofr}
\end{equation}
and the function $r(k)$ is in $C^2(\RR)$ and decays as $O(1/|k|^3)$ as $|k|\to \infty$. The time evolution of the quantities $r(k,t)$ and $\ga_n(t)$ is given by
\begin{equation}
r(k,t)=r(k)e^{8ik^3t},\quad \ga_j(t)=\ga_n e^{4\ka_n^3 t}.
\end{equation}

\end{proposition}

The collection $\left(r(k,t), k\geq 0; \ka_1,\ldots,\ka_N,\ga_1(t),\ldots,\ga_N(t)\right)$ is called the scattering data of the Schr\"odinger operator $L(t)$. We encode the scattering data as a contour problem in the following way. Consider the function
\begin{equation}
\chi(k,x,t)=\left\{\begin{array}{cc} t(k) \psi_-(k,x,t)e^{ikx}, & \Im k>0, \\ \psi_+(-k,x,t)e^{ikx}, & \Im k<0.\end{array}\right.\label{eq:chi}
\end{equation}

\begin{proposition}[See Thm.~2.3 in \cite{GT09}] Let $(r(k); \ka_1,\ldots,\ka_N,\ga_1,\ldots,\ga_N)$ be the scattering data of the Schr\"odinger operator $L(0)$. Then the function $\chi(k,x,t)$ defined by \eqref{eq:chi} is the unique function satisfying the following properties: \label{prop:ISM}

\begin{enumerate}

\item $\chi$ is meromorphic on the complex $k$-plane away from the real axis and has non-tangential limits \label{item:meromorphic}
\begin{equation}
\chi_{\pm}(k,x,t)=\lim_{\varepsilon \to 0}\chi(k\pm i\varepsilon,x,t),\quad k\in \RR. \label{eq:limits}
\end{equation}
on the real axis.

\item $\chi$ has a jump on the real axis satisfying
\begin{equation}
\chi_+(k,x,t)-\chi_-(k,x,t)=r(k)e^{2ikx+8ik^3t} \chi_-(-k,x,t),\quad k\in \RR. \label{eq:chijumponrealaxis}
\end{equation}

\item $\chi$ has simple poles at the points $i\ka_1,\ldots,i\ka_N$ and no other singularities. The residues at the poles satisfy the condition
\begin{equation}
\Res_{i\ka_n}\chi(k,x,t)=ic_n e^{-2\ka_nx+8\ka_n^3t}\chi(-i\ka_n,x,t),\quad c_n=\ga_n^2.
\label{eq:residues}
\end{equation}

\item $\chi$ has the asymptotic behavior \label{item:asymp}
\begin{equation}
\chi(k,x,t)=1+\frac{i}{2k}Q_+(x,t)+O\left(\frac{1}{k^2}\right),\quad |k|\to \infty,\quad \Im k\neq 0.
\label{eq:asymptotic1}
\end{equation}
\end{enumerate}
The function $\chi$ is a solution of the equation
\begin{equation}
\chi''-2ik\chi'-u(x)\chi'=0, \label{eq:chidiff}
\end{equation}
and the function $u(x,t)$ given by the formula
\begin{equation}
u(x,t)=\dfrac{d}{dx}Q_+(x,t)\label{eq:uintermsofQ}
\end{equation}
is a solution of the KdV equation \eqref{eq:KdV} satisfying condition \eqref{eq:udecay}.
\end{proposition}

We now reformulate, following Manakov and Zakharov~\cite{MZ85}, the conditions~\eqref{eq:limits}-\eqref{eq:asymptotic1} that define $\chi$ as an $\opa$-problem. Denote by $\rho(k,x,t)$ the jump of $\chi$ on the real axis, going from the positive to the negative side, and denote $i\chi_n(x,t)$ the residue of $\chi(k,x,t)$ at $k=i\ka_n$. Conditions~\eqref{eq:limits}-\eqref{eq:asymptotic1} imply that $\chi$ has a spectral representation
\begin{equation}
\chi(k,x,t)=1+\frac{1}{2\pi i}\int_{-\infty}^{\infty}\frac{\rho(q,x,t)}{q-k} dq+i\sum_{n=1}^N \frac{\chi_n(x,t)}{k-i\ka_n},
\end{equation}
and that the jump $\rho(k,x,t)$ and the residues $\chi_n(x,t)$ satisfy the system
\begin{equation}
\rho(k,x,t)=r(k)e^{2ikx+8ik^3t}\left[1+\frac{1}{2\pi i} \int_{-\infty}^{\infty} \frac{\rho(q,x,t)}{q+k+i\varepsilon}dq-i\sum_{n=1}^N \frac{\chi_n(x,t)}{k+i\ka_n}\right],
\end{equation}
\begin{equation}
\chi_n(x,t)=c_ne^{-2\ka_nx+8\ka_n^3t}\left[1+\frac{1}{2\pi i}\int_{-\infty}^{\infty} \frac{\rho(q,x,t)}{q+i\ka_n}dq+\sum_{m=1}^N\frac{\chi_m(x,t)}{\ka_n+\ka_m} \right].
\end{equation}
We now rewrite, following~\cite{MZ85}) the conditions~\eqref{eq:limits}-\eqref{eq:asymptotic1} that define $\chi$ (see~\cite{MZ85}) as a $\opa$-problem. Consider the distribution 
\begin{equation}
T(k)=\frac{i}{2}\delta(k_I)\theta(-k_I) r(k_R) + \pi \delta(k_R)\sum_{n=1}^N c_n \delta(k_I-\ka_n),
\label{eq:T}
\end{equation}
called the {\it dressing function}. Here $\theta$ is the Heaviside step function, $\delta$ is the Dirac delta function, $k=k_R+ik_I$, and we use that
\begin{equation}
\frac{\pa}{\pa \ok}\frac{1}{k}=\pi\delta(k)=\pi\delta(k_R)\delta(k_I).
\end{equation}
The meaning of the distributions $\delta(x)\theta(\pm x)$ is that, for a function $f$ that is possibly discontinuous at $x=0$, we have
\begin{equation}
\int_{-\infty}^{\infty} f(x)\delta(x)\theta (\pm x)dx=\lim_{x\to 0^{\pm}} f(x).
\end{equation}
A direct calculation then shows that conditions~\eqref{eq:chijumponrealaxis}-\eqref{eq:residues} are equivalent to the following $\opa$-problem (see~\cite{MZ85}):
\begin{equation}
\frac{\pa\chi}{\pa \ok}=T(k)e^{2ikx+8ik^3t} \chi(-k,x,t),\quad \chi\to 1\mbox{ as }k\to\infty.
\label{eq:opaproblem}
\end{equation}

\section{Transplanting poles} \label{sec:transplanting}

The key observation of~\cite{DZZ16,ZDZ16,ZZD16} is that, in order to obtain finite-gap solutions of KdV from $N$-soliton solutions in the $N\to\infty$ limit, it is necessary to first allow $\chi$ to have poles on the negative imaginary axis. We now perform the same procedure for a generic rapidly decreasing potential.

Let $(r(k),\ka_1,\ldots,\ka_N,c_1,\ldots,c_N)$ be the scattering data of a potential $u(x,t)$ rapidly decreasing at infinity, and let $\chi(k,x,t)$ be the function determined by Prop.~\ref{prop:ISM}. Fix a subset $I\subset \{1,\ldots,N\}$, and introduce the function
\begin{equation}
\tchi(k,x,t)=\chi(k,x,t)\prod_{m\in I}\frac{k-i\ka_m}{k+i\ka_m}.
\end{equation}
The function $\tchi$ has a jump on the real axis, tends to $1$ as $k\to\infty$, has $N-\#I$ poles in the upper half-plane at $k=i\ka_m$ for $m\notin I$, and $\# I$ poles in the lower half-plane at $k=-i\ka_m$ for $m\in I$. It follows that $\tchi$ has the spectral representation
\begin{equation}
\tchi(k,x,t)=1+\frac{1}{2\pi i}\int_{-\infty}^{\infty}\frac{\trho(q,x,t)}{q-k}dq+i\sum_{m\notin I}\frac{\tchi_m(x,t)}{k-i\ka_m}+i\sum_{n\in I}\frac{\tchi_m(x,t)}{k+i\ka_m},
\end{equation}
where the jump $\trho$ and the residues $i\tchi_n$ are equal to 
\begin{equation}
\trho(q,x,t)=\rho(q,x,t)\prod_{m\in I}\frac{q-i\ka_m}{q+i\ka_m},\quad \tchi_n(x,t)=\chi_n(x,t)\prod_{m\in I}\frac{\ka_n-\ka_m}{\ka_n+\ka_m}\mbox{ for }n\notin I,\end{equation}
\begin{equation}
\tchi_n(x,t)=-2\ka_n\chi(-i\ka_n,x,t)\prod_{m\in I\backslash\{n\}}\frac{\ka_n+\ka_m}{\ka_n-\ka_m}\mbox{ for }n\in I.
\end{equation}
It follows that $\tchi$ satisfies same $\opa$-problem~\eqref{eq:opaproblem} as $\chi$, but with the dressing function
\begin{equation}
\tT(k)=\frac{i}{2}\delta(k_I)\theta(-k_I) \tr(k_R) + \pi \delta(k_R)\sum_{n=1}^N \tc_n \delta(k_I-\tka_n)
\label{eq:tT}
\end{equation}
whose coefficients are equal to 
\begin{equation}
\tr(k)=r(k)\prod_{m\in I}\left(\frac{k-i\ka_m}{k+i\ka_m}\right)^2,\quad \tc_n=c_n\prod_{m\in I}\left(\frac{\ka_n-\ka_m}{\ka_n+\ka_m}\right)^2\mbox{ for }n\notin I,
\end{equation}
\begin{equation}
\tc_n=-\frac{4\ka_n^2}{c_n}\prod_{m\in I\backslash\{n\}}\left(\frac{\ka_n+\ka_m}{\ka_n-\ka_m}\right)^2\mbox{ for }n\in I,\quad \tka_n=\left\{\begin{array}{cc}\ka_n, & n\notin I,\\ -\ka_n, & n\in I.\end{array}\right.\label{eq:transplantedcoefficients}
\end{equation}
We observe that
\begin{equation}
\tr(-k)=\overline{\tr(k)},\quad |\tr(k)|=|r(k)|\mbox{ for }k\in \RR,\quad \tr(0)=r(0),
\end{equation}
hence the function $\tr$ satisfies the same properties~\eqref{eq:symmetryofreflection}-\eqref{eq:propertiesofr} as $r$. We also note that $\tc_n$ is positive if $n\notin I$ and negative if $n\in I$, in other words $\tc_n$ has the same sign as $\tka_n$.

Finally, we observe that $\tchi$ satisfies the same differential equation~\eqref{eq:chidiff} as $\chi$, and has the following asymptotic behavior as $|k|\to\infty$:
\begin{equation}
\tchi(k,x,t)=1+\frac{i}{2k}\tQ_+(x,t)+O\left(\frac{1}{k^2}\right),\quad \tQ_+(x,t)=Q_+(x,t)-4\sum_{m\in I}\ka_m.
\end{equation}
Therefore $u(x,t)$ is obtained from $\tchi(k,x,t)$ using the same formula~\eqref{eq:uintermsofQ}.

We have shown that if $u(x,t)$ is obtained by the ISM from the spectral data $(r(k),\ka_1,\ldots,\ka_N,c_1,\ldots,c_N)$, then we can change the sign of the $\ka_m$ for $m\in I$, modify the coefficients $r$ and $c_n$ according to~\eqref{eq:transplantedcoefficients}, and the resulting function $\tchi$ produces the same potential $u(x,t)$. The result of the above procedure is that we have transplanted some of the poles of $\chi$ to the negative imaginary axis without changing $u$.

Reversing this procedure, we claim the following. Let $(r(k);\ka_1,\ldots,\ka_N,c_1,\ldots,c_N)$ be data consisting of a function $r:\RR\to \RR$ satisfying properties~\eqref{eq:symmetryofreflection}-\eqref{eq:propertiesofr} and $2N$ nonzero real constants $\ka_n$, $c_n$ such that
\begin{equation}
\ka_n\neq \pm \ka_m\mbox{ for any }n\neq m,\quad \ka_n/c_n>0\mbox{ for all }n. \label{eq:transplantedc}
\end{equation}
Then the $\opa$-problem~\eqref{eq:opaproblem} with the dressing function $T$ given by~\eqref{eq:T} has a unique solution $\chi$, and the function $u(x,t)$ given by~\eqref{eq:uintermsofQ} is a solution of KdV with $N$ solitons having spectral parameters $|\ka_1|,\ldots,|\ka_N|$. Therefore, every solution of KdV with $N$ solitons (and a non-trivial reflection coefficient) can be obtained using the dressing method in $2^N$ different ways.

\section{Construction of generalized primitive potentials}\label{sec:generalized}

We now pass to the $N\to\infty$ limit, as in~\cite{DZZ16,ZDZ16,ZZD16}. Consider the $\opa$-problem~\eqref{eq:opaproblem} with a dressing function $T$ of the form~\eqref{eq:T}, where $r$ satisfies 
\eqref{eq:symmetryofreflection}-\eqref{eq:propertiesofr} and the $\ka_n$ and $c_n$ satisfy~\eqref{eq:transplantedc}. Fix $0<k_1<k_2$, and suppose that the poles $\ka_n$ are uniformly distributed in the two intervals $[k_1,k_2]$ and $[-k_2,-k_1]$, in such a way that $\ka_n\neq \pm \ka_m$ for $n\neq m$. In the limit as $N\to \infty$, we obtain the following dressing function:
\begin{equation}
T(k)=\frac{i}{2}\delta(k_I)\theta(-k_I)r(k)+\pi \de(k_R)\left[\int_{k_1}^{k_2}R_1(p)\de(k_I-p)dp-\int_{k_1}^{k_2} R_2(p) \de(k_I+p)dp\right].
\label{eq:newT}
\end{equation}
We impose the same conditions on $r(k)$ as before:
\begin{equation}
r(-k)=\overline{r(k)},\quad |r(k)|<1\mbox{ for }k\neq 0,\quad |r(0)|\leq 1,\quad r(0)=-1\mbox{ if }|r(0)|=1.
\end{equation}
We require the functions $R_1$ and $R_2$ on $[k_1,k_2]$ to be positive and H\"older-continuous. 

We now consider a solution $\chi$ of the $\opa$-problem~\eqref{eq:opaproblem} with a dressing function of the form~\eqref{eq:newT}. Such a $\chi$ has a jump on the real axis as well as on the intervals $[ik_1,ik_2]$ and $[-ik_2,-ik_1]$ on the imaginary axis, and has the following spectral representation:
\begin{equation}
\chi(k,x,t)=1+\frac{1}{2\pi i}\int_{-\infty}^{\infty}\frac{\rho(p,x,t)dp}{p-k}+i\int_{k_1}^{k_2}\frac{f(p,x,t)dp}{k-ip}+i\int_{k_1}^{k_2}\frac{g(p,x,t)dp}{k+ip}.
\end{equation}
Denote $\chi^{\pm}(ik,x,t)$ the right and left limit values of $\chi$ on the cuts $k\in [ik_1,ik_2]$ and $k\in[-ik_1,-ik_2]$ on the imaginary axis, then
\begin{equation}
\chi^+(ik,x,t)-\chi^-(ik,x,t)=2\pi i f(p,x,t),\quad \chi^+(ik,x,t)-\chi^-(-ik,x,t)=2\pi ig(k,x,t).
\end{equation}

On the real axis, $\chi$ satisfies the Riemann--Hilbert problem~\eqref{eq:chijumponrealaxis}, where $\chi^{\pm}$ are the upper and lower limits of $\chi$ on the real axis. On the cuts $k\in[ik_1,ik_2]$ and $k\in[-ik_2,-ik_1]$ on the imaginary axis, the function $\chi$ satisfies the symmetric Riemann--Hilbert problem
\begin{equation}
f(k,x,t)=\frac{1}{2}R_1(k)e^{-2kx+8k^3t}\left[\chi^+(-ik,x,t)+\chi^-(-ik,x,t)\right],\label{eq:imaginaryjump1}
\end{equation}
\begin{equation}
g(k,x,t)=-\frac{1}{2}R_2(k)e^{2kx-8k^3t}\left[\chi^+(ik,x,t)+\chi^-(ik,x,t)\right].\label{eq:imaginaryjump2}
\end{equation}

Together, the Riemann--Hilbert problems~\eqref{eq:chijumponrealaxis},~\eqref{eq:imaginaryjump1},~\eqref{eq:imaginaryjump2} are equivalent to the following system of singular integral equations on $\rho$, $f$, and $g$:
\begin{equation}
\rho(k,x,t)=r(k,x,t)e^{-2ikx-8ik^3t}\left[1+\frac{1}{2\pi i}\int_{-\infty}^{\infty}\frac{\rho(p,x,t)dp}{q+ik-\varepsilon}-i\int_{k_1}^{k_2}\frac{f(p,x,t)dp}{k+ip}+i\int_{k_1}^{k_2}\frac{g(p,x,t)dp}{-k+ip}\right],\label{eq:rho2}
\end{equation}
\begin{equation*}
f(k,x,t)+R_1(k)e^{-2kx+8k^3t}\left[\int_{k_1}^{k_2}\frac{f(p,x,t)}{k+p}dp+\fint_{k_1}^{k_2}\frac{g(p,x,t)}{k-p}dp \right]=
\end{equation*}
\begin{equation}
=R_1(k)e^{-2kx+8k^3t}\left[1+\frac{1}{2\pi i}\int_{-\infty}^{\infty} \frac{\rho(p,x,t)}{p-ik}dp\right],\label{eq:RH1}
\end{equation}
\begin{equation*}
g(k,x,t)+R_2(k)e^{2kx-8k^3t}\left[\fint_{k_1}^{k_2}\frac{f(p,x,t)}{k-p}dp+\int_{k_1}^{k_2}\frac{g(p,x,t)}{k+p}dp \right]=
\end{equation*}
\begin{equation}
=-R_2(k)e^{2kx-8k^3t}\left[1+\frac{1}{2\pi i}\int_{-\infty}^{\infty} \frac{\rho(p,x,t)}{p+ik}dp\right].\label{eq:RH2}
\end{equation}
The system of equations~\eqref{eq:rho2}, \eqref{eq:RH1}, \eqref{eq:RH2} is a complete system of equations defining generalized primitive potentials (for fixed $t$) and corresponding solutions of KdV. Setting $r(k)=0$ yields $\rho(k,x,t)=0$, and we obtain the system of equations describing reflectionless primitive potentials that we derived in our previous papers \cite{DZZ16,ZDZ16,ZZD16,ZZ18}.

The corresponding solution $u(x,t)$ of KdV is  given by the formula
\begin{equation}
u(x,t)=2\frac{d}{dx}\left[-\frac{1}{2\pi} \int_{-\infty}^{\infty}\rho(p,x,t)dp+\int_{k_1}^{k_2}[f(p,x,t)+g(p,x,t)] dp\right].
\end{equation}

\section{Conclusion}

We stress that the class of generalized primitive potentials that we described is much broader than the reflectionless primitive potentials that we studied in our previous work. They inherit some of their properties. The have the same spectral set $[-k_2^2,-k_1^2]\cup [0,\infty)$, and the spectrum on the real positive axis $(0,\infty)$ is also doubly degenerate. However, these potentials are no longer reflectionless on $[0,\infty)$, and the corresponding $\chi$-function has a jump on the real axis. We formulate two hypotheses that we hope to prove in the future:

\begin{enumerate} \item We conjecture that we have described all bounded potentials of the Schr\"odinger operator having the spectral set $[-k_2^2,-k_1^2]\cup [0,\infty)$ that are doubly degenerate on the positive real axis. We note that this limitation is very strict. We note that this condition coincides with the requirement that $|c(k)|<1$ for $k\neq 0$. Potentials having $|c(k)|=1$ on some subset of the real axis certainly do exist, but they have not been properly described so far.

\item We believe that the generalized primitive potentials that we have described are spatially localized perturbations of the reflectionless primitive potentials. Here we refer to the paper of Kuznetsov and Mikhailov \cite{KM77}, who studied localized perturbations of the simplest primitive potential, namely the one-gap periodic potential expressed in terms of the Weierstrass $\wp$-function. 

\end{enumerate}

\section{Funding}

The first author gratefully acknowledges the support of NSF grant DMS-1716822. The second author gratefully acknowledges the support of NSF grant DMS-1715323. The results in Sections~\ref{sec:IST}-\ref{sec:generalized} were obtained with the support of Grant RScF 19-72-30028.

\end{document}